\documentclass[manuscript,screen,nonacm]{acmart}
\usepackage{paralist}
\usepackage{enumerate}

\usepackage{microtype}
\usepackage{cleveref}

\usepackage{tikz}
\usetikzlibrary{decorations.pathmorphing}

\usepackage{color, colortbl}
\definecolor{LightGray}{gray}{0.9}

\usepackage{paralist}
\usepackage{enumerate}

\usepackage{microtype}
\usepackage{bm}
\usepackage{cleveref}

\usepackage{tikz}
\usetikzlibrary{decorations.pathmorphing} 

\usepackage{pifont}
\usepackage[utf8]{inputenc}
\usepackage[T1]{fontenc}
\usepackage{newunicodechar}
\usepackage{threeparttable}

\newunicodechar{❌}{\ding{55}} % From pifont
\newunicodechar{✔}{\ding{51}} % Approximate
\newunicodechar{⚫}{\ding{108}}
\newunicodechar{〇}{\ding{109}} 

\newcommand{\ntia}[1]{}
\AtBeginDocument{%
  }

\setcopyright{acmlicensed}
\copyrightyear{2018}
\acmYear{2018}
\acmDOI{XXXXXXX.XXXXXXX}
\acmConference[Conference acronym 'XX]{Make sure to enter the correct
  conference title from your rights confirmation email}{June 03--05,
  2018}{Woodstock, NY}
\acmISBN{978-1-4503-XXXX-X/2018/06}

\begin{document}

\title{Towards AI Transparency and Accountability: A Global Framework for Exchanging Information on AI Systems}

    \author{Warren Buckley}
    \authornote{Both authors contributed equally to this paper.}
    \affiliation{%
      \institution{University College Dublin}
      \city{Dublin}
      \country{Ireland}}
    \email{patrick.w.buckley@ucdconnect.ie}

\author{Adrian Byrne}
\affiliation{%
  \institution{University College Dublin}
  \city{Dublin}
  \country{Ireland}}
\email{adrian.byrne@ucd.ie}

\author{Nicholas Perello}
\affiliation{%
  \institution{University of Massachusetts Amherst}
  \city{Amherst}
  \state{Massachusetts}
  \country{USA}}
\email{nperello@umass.edu}

\author{Cyrus Cousins}
\affiliation{%
  \institution{Duke University}
  \city{Durham}
  \state{North Carolina}
  \country{USA}}
\email{originalcyruscousins@gmail.com}

\author{Taha Yasseri}
\affiliation{%
  \institution{Trinity College Dublin, Technological University Dublin}
  \country{Ireland}}
\email{Taha.Yasseri@tcd.ie}

\author{Yair Zick}
\affiliation{%
  \institution{University of Massachusetts Amherst}
  \city{Amherst}
  \state{Massachusetts}
  \country{USA}}
\email{yzick@umass.edu}

\author{Przemyslaw Grabowicz}
\authornotemark[1]
\affiliation{%
  \institution{University College Dublin}
  \city{Dublin}
  \country{Ireland}}
\email{przemek.grabowicz@ucd.ie}

\renewcommand{\shortauthors}{Buckley et al.}

\begin{abstract}
  
We propose that future AI transparency and accountability regulations are based on an open global standard for exchanging information about AI systems, which allows co-existence of potentially conflicting local regulations. Then, we discuss key components of a lightweight and effective AI transparency and/or accountability regulation. To prevent overregulation, the proposed approach encourages collaboration between regulators and industry to create a scalable and cost-efficient mutually beneficial solution. This includes using automated assessments and benchmarks with results transparently communicated through AI cards in an open AI register to facilitate meaningful public comparisons of competing AI systems. Such AI cards should report standardized measures tailored to the specific high-risk applications of AI systems and could be used for conformity assessments under AI transparency and accountability policies such as the European Union's AI Act.

\end{abstract}



\maketitle

%\textit{This white paper is a response to the ``\href{https://www.federalregister.gov/documents/2023/04/13/2023-07776/ai-accountability-policy-request-for-comment}{AI Accountability Policy Request for Comments}'' by the National Telecommunications and Information Administration of the United States. The question numbers for which comments were requested are provided in superscripts at the end of key sentences answering the respective questions. The white paper offers a set of interconnected recommendations for an AI accountability policy.}

\section{Introduction}
As new AI agents and models are released and claims of their novel capabilities are made, it is crucial for governments, corporations, and individual customers to verify their capabilities, safety, impartiality, and limitations. AI systems often align more with the interests of their creators and maintainers and vary in terms of their political, corporate, and national perspectives and biases. 
How can countries and their governments incentivize AI system designs that best fulfill the needs, interests, and safety of the public while supporting AI innovation in a world of varying geopolitical AI stances?

This is a crucial question in times when multiple jurisdictions develop their AI transparency or accountability policies. The most advanced AI regulation to date has been developed and adopted -- arguably -- by the European Union (EU). The EU adopted the AI Act in 2024, establishing a risk-based framework for AI accountability and transparency, aimed at ensuring a high level of protection of health, safety, and fundamental rights.\footnote{https://eur-lex.europa.eu/eli/reg/2024/1689/oj/eng} However, details of the implementation of the act are not determined yet. The European Commission (EC) over the next years plans to develop a series of delegated and implementing acts detailing the policy in communication with AI experts and the EU public.\footnote{{https://artificialintelligenceact.eu/implementation-timeline}} Other jurisdictions globally explore the possibility of introducing similar AI policies. Here, we propose and discuss (i) a framework that brings potentially conflicting -- and developed at different paces -- AI policies under one umbrella and (ii) key components of AI transparency and accountability regulations. To contextualize these points, throughout the manuscript, we use the AI Act as an example of an advanced and developing AI regulation, and as a reference point that is different -- yet compatible -- to our proposal.
% \footnote{For instance, in 2023, the U.S. administration requested public comments on the AI Accountability Policy to be submitted to National Telecommunications and Information Administration.}

The EU AI Act has faced immense resistance from industry and from other governments,\footnote{https://www.theguardian.com/us-news/2025/aug/26/donald-trump-tariffs-us-tech-uk-digital-services-tax-eu} including calls to pause its roll out over concerns it could stifle innovation. In response, the EC sought submissions in late 2025 on the simplification of the EU AI Act rollout.\footnote{https://oeil.europarl.europa.eu/oeil/en/procedure-file?reference=2025/0359(COD)} The submissions to date reflect the complexity of regulating AI. On the one hand, Meta Platforms (the parent company of Facebook) asks to "pause the implementation".\footnote{https://ec.europa.eu/info/law/better-regulation/have-your-say/initiatives/14855-Simplification-digital-package-and-omnibus/F33089184\_en} On the other hand, the Ada Lovelace Institute appeals that the simplifications should not be at the "expense of people, society or fundamental rights".\footnote{https://ec.europa.eu/info/law/better-regulation/have-your-say/initiatives/14855-Simplification-digital-package-and-omnibus/F33089199\_en} Other stakeholders, e.g., OpenAI, recognize the need to protect people but request a much simplified regulation.\footnote{https://ec.europa.eu/info/law/better-regulation/have-your-say/initiatives/14855-Simplification-digital-package-and-omnibus/F33088837\_en} Policies and regulations take time to draft and implement, often lagging behind the pace of developments in industry that they seek to regulate. In AI, this pace mismatch threatens the effectiveness of regulations and promotes the view that regulation is costly, stifles innovation and runs counter to market-driven product innovation.  While industry is concerned with the burden of overregulation, it does speak of the need for responsibility. For example, the CEO of Google DeepMind, Demis Hassabis, acknowledges the urgent need for international collaboration between institutions and industry dealing with AI's impact on society.\footnote{https://www.youtube.com/watch?v=PqVbypvxDto} Similarly, researchers emphasize the importance of understanding ``machine behavior''~\cite{rahwan2019machine}.

\begin{figure}
     \vspace{-0.3cm}
     \centering
     \includegraphics[width=0.99\linewidth]{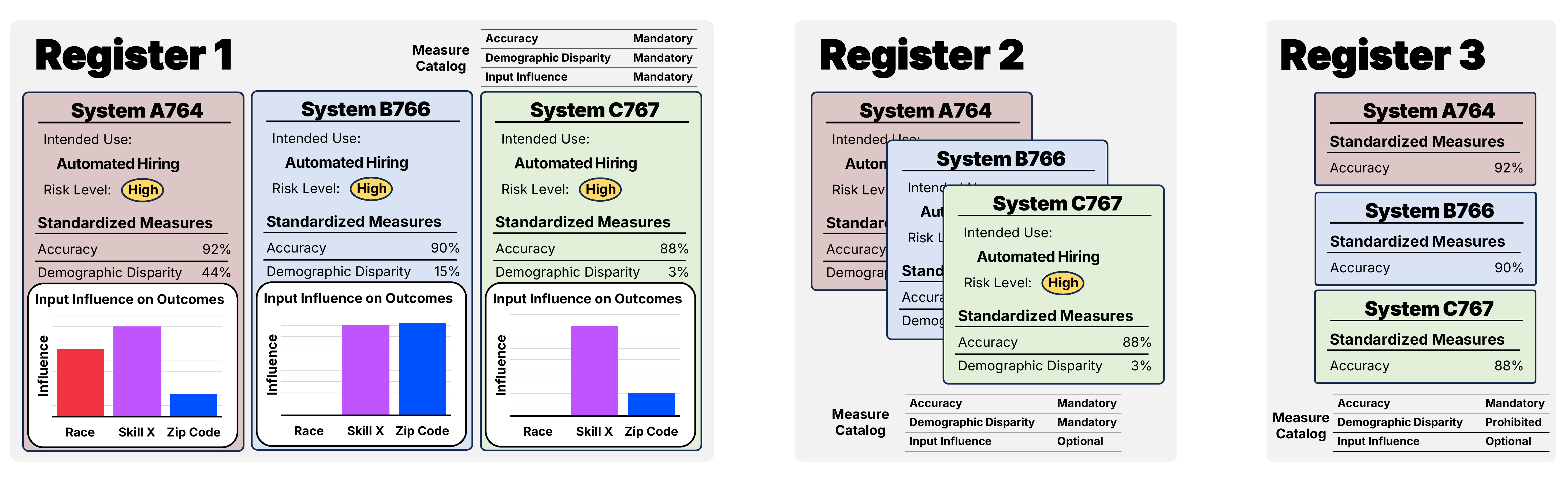}
     \caption{Example entries in three hypothetical AI registers for AI systems whose intended use is hiring automation. The listed standardized measures are exemplary -- respective AI Offices shall determine appropriate Measures (possibly specific to intended use of AI systems). The same system may be in multiple registers, sharing System IDs but complying with local requirements, e.g., regarding the disclosed Measures. Here, comparisons of input influence Measures in Register 1 suggest that System A764 is directly discriminatory with respect to race, while System B766 is indirectly discriminatory via a proxy of race (Zip code).}
     \label{fig:comparison_card}
\end{figure}
Acknowledging \textbf{(i)} the urgent need for transparency, accountability and control, \textbf{(ii)} the friction between policy-makers and industry on the degree of regulation, and \textbf{(iii)} the global diversity in AI policy stances, we propose a global federated framework for the capture and exchange of information about AI systems to address all three challenges (outlined in \S\ref{sec:anatomy}). We believe such a framework, centered around transparency, will facilitate the capture and comparison of key information about the identity of AI legal entities and their AI systems, their regulatory status in jurisdictions and the results of various AI measures. We catalog existing jurisdictional regulations and then outline the benefits, to AI policy-makers, the AI industry and the public, of putting AI transparency first by supporting heterogeneous policies on a unified framework (\S\ref{sec:alignment}). We acknowledge that each jurisdictions can make given AI measures mandatory or optional, such as the ones illustrated in Figure~\ref{fig:comparison_card}, and can control the level of access the public has to information. We present that it is in the long-term interest of the AI industry to work with research bodies to develop AI assessments and measures that are of interest to the public and that enables them to make informed decisions, while reducing the need for all-encompassing regulation (\S\ref{sec:ai-measures}). 

%TODO NP: What we don't do and reinforce novelty above. And reword below to be "Our Contributions"
%federated identifiers + standardized measures + automated assessments under a global technical foundation
%\subsection{Our Contributions}
%Section (\S\ref{sec:anatomy}) outlines the key components of the proposed federated global framework, Section (\S\ref{sec:alignment}) catalogs existing jurisdictional regulations and outlines the benefits accrued to AI policy-makers, the AI industry and the public, by putting AI transparency first by supporting heterogeneous policies on a unified framework and Section (\S\ref{sec:ai-measures}) outlines through example the need for standardized AI system measures available for public comparison.

% Such a framework would instill long-term and broad international trust in techno-social systems by incentivizing trustworthy AI designs, through increased transparency and minimal viable regulation. %\ntia{1}  
% In this context, we define accountability as encompassing: (1) the ability to characterize a system and meaningfully compare it with other systems, (2) clear reasoning behind its outcomes, e.g., in terms of its inputs, and
% (3) the ability to take action when things go wrong. These pillars frame the technical and regulatory mechanisms we propose. 

\section{Anatomy of the Proposed AI Information Framework}
\label{sec:anatomy}
\begin{figure}
    \centering
    \includegraphics[width=0.90\linewidth]{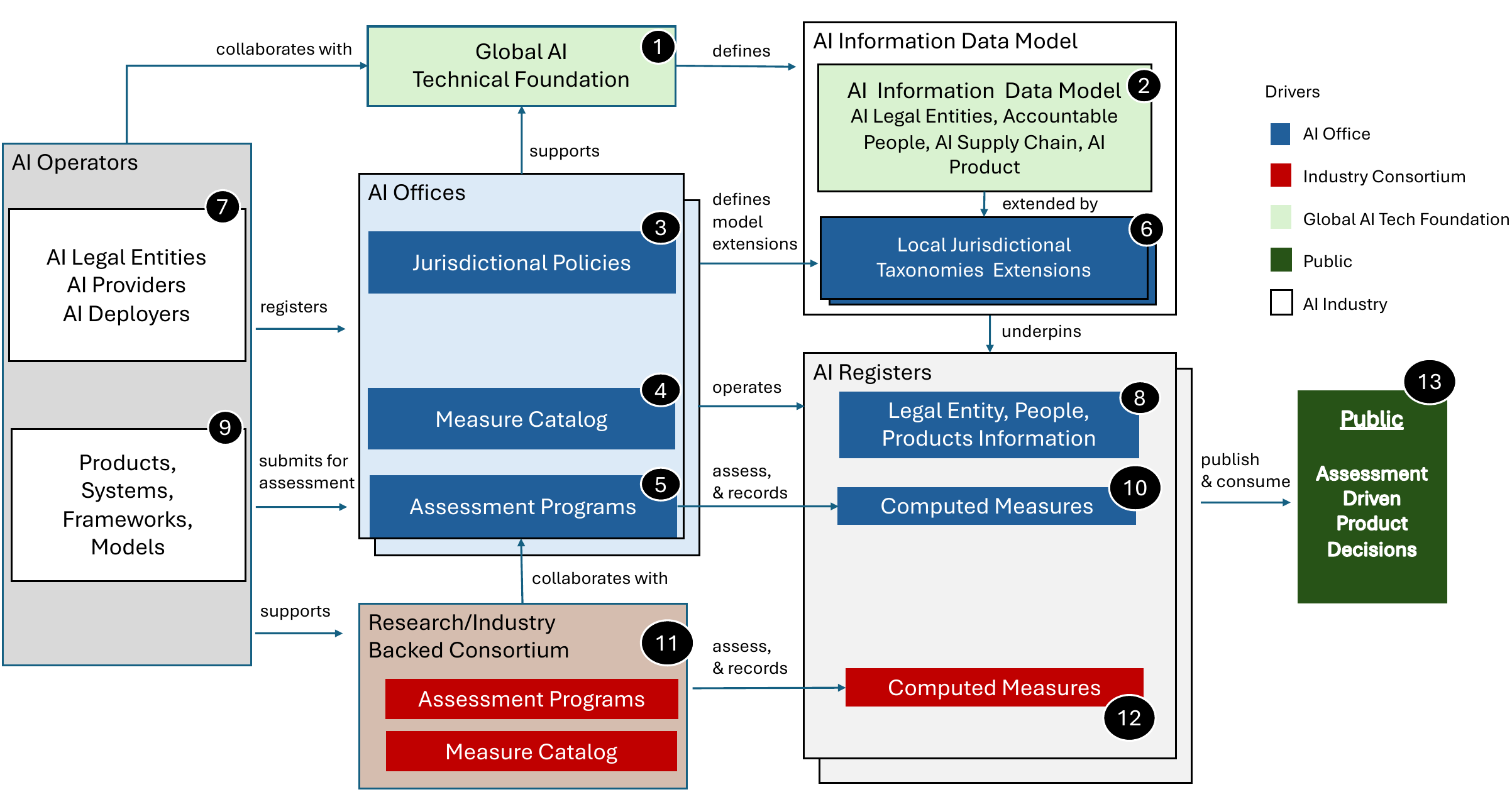}
    \caption{Global Framework for Exchanging Information on AI Systems. A Global AI Technical Foundation (1) creates an extensible AI data model (2). Governmental policy-makers in an AI Office create jurisdiction specific policies (3), Measure Catalog (4) and official Assessment Programs (5). They enhance the data model with jurisdictional taxonomies (6). AI Operators (7) comply with government regulations to register key reference data in AI Register (8). AI systems (9) are evaluated via an AI Office's Assessment Programs (5) and the resulting computed Measures (10) are stored in an AI Register. Industry Backed Assessment Programs (11) can mirror AI Offices' Measure Catalog and AI Offices' Assessment Programs as a baseline to produce new measures and assessments. The public (13) can make informed comparisons and decisions on their direct and indirect use of AI.}
\label{fig:framework-overview}
\vspace{-0.3cm}    
\end{figure}

The proposed framework (see Figure~\ref{fig:framework-overview}) includes: an \textbf{AI Information Data Model} to capture key information about AI systems;
robust \textbf{Identifier Allocation} for key AI data model artifacts;
a \textbf{Catalog of Standardized Measures} that can be computed consistently when evaluating AI systems in the context of a given intended use;
\textbf{Assessment Programs} as a means for efficient and scalable AI conformity assessment including automated assessments where possible;
\textbf{AI Registers} for the storage and distribution of AI system information;
\textbf{AI Offices} for the implementation of the jurisdiction specific administrative and regulatory requirements;
and a \textbf{Global AI Technical Foundation} to oversee the AI framework independent of any jurisdiction's policies.  AI Offices customize Assessments Programs, manage the Catalog of Standardize Measures and define the AI systems to be covered based on their policies - e.g., the EU AI Act has a very specific AI systems definition within Article 57.

At first, it may seem to be a complex unachievable goal to build such a framework but we draw from other similar initiatives that have been successful in domains such as finance and car safety. We outline an approach that can underpin the range of goals and pace of execution of different AI policy-makers while reducing the burden on the AI industry by encouraging standardization, and the reduction of duplication through the reuse of key artifacts such as AI evaluation measures and AI system information. As the EU AI Act seeks to crystallize its technical implementation we would encourage it to adopt such a framework and encourage collaboration with industry and other jurisdictions to spearhead the foundational elements of such a framework. The components of the framework are described below.

\subsection{AI Information Data Model, Standardized Identifiers and Global Technical Foundation}
\textbf{AI Information Data Model}: A data model which would support the capture of information about AI Operators, which are defined in the EU's AI Act, as well as their AI products and AI deployments. This would enable and encourage traceability and accountability of AI technology from ultimate source, through enterprise supply chains, right through to consumers and the impacted public. This is important as we see multi-layer AI architectures including cascading agentic AI systems composed of many AI components. The level of detail required to be captured would be determined by respective regulation of an AI Office. The data model can leverage research on modeling AI Use Policies and AI Use Cases~\cite{hupont2024cards,golpayegani2024ordl}.

\textbf{Identifier Allocation}: Identifiers are key to any data model. We propose a federated Identifier Allocation system so that AI Operators, as well as their AI systems, would be allocated unique identifiers by the framework. We propose that the identification codes for products and models be administered similarly to those used by the Global Legal Entity Identifier Foundation (GLEIF)\footnote{https://www.gleif.org/en and https://lei-ireland.ie/ and https://www.fsb.org/about/rcgs/} which delegates the Legal Entity Identifiers (LEI) governance to local jurisdictions.  

\textbf{Jurisdiction Differences and Data Model Extensibility}: While rigid data models are difficult to agree globally, this problem can be solved using extensible models, which are common in financial data modeling and cater to jurisdiction variations. For example, the eXtensible Business Reporting Language (XBRL) a system for company filings allows for variations by jurisdiction\footnote{https://www.xbrl.org/the-consortium/about/jurisdictions/} while maintaining a common underlying data model. 
%For example, the specific measures and information available to the public would be determined at a jurisdiction level similar to how US XBRL is published and controlled by the SEC's EDGAR database\footnote{https://www.sec.gov/data-research/structured-data/inline-xbrl} independent of other jurisdictions yet sharing common data model elements.  

\textbf{Global AI Technical Foundation}: We envisage a Global AI Technical Foundation (GAITF) to oversee the AI Information Data Model and the Identifier Allocation system. The foundation can leverage the approaches similar to the globally accepted LEI and XBRL frameworks mentioned above or equivalent. We believe centralizing  technical modeling in the GAITF, which services decentralized AI Offices, meets the requirements of simplifying interactions for the AI industry while allowing local jurisdictional variations.   The foundation would be supported, funded and governed by AI Offices and would collaborate with the AI industry . 

%information exchange framework and foundation data model. Similar to XBRL the model would allow taxonomies to be used to expand the model without breaking the foundation model, this will allow for jurisdiction specific AI regulations similar to XBRL's local financial filing taxonomies.

\subsection{AI Offices and AI Registers}
\textbf{AI Offices:} While the GAITF would look after the technical framework, jurisdiction specific AI Offices would be established to look after their own specific policy and information requirements. In addition, these offices would leverage the data model provided by the GAITF. The offices would determine elements of the model that are mandatory or need to be extended at the level appropriate for their jurisdiction. AI Offices can in turn further delegate to other offices as required, e.g., the EU AI Office\footnote{\url{https://digital-strategy.ec.europa.eu/en/policies/ai-office}} could be an EU-level AI Office with delegation to national AI Offices as required.

\textbf{AI Registers:} AI Offices would implement AI Registers aligned with the AI Information Data Model. As AI Operators register with AI Offices, identifiers would be allocated (for sharing with other offices) and system information would be stored. Key information can be shared across AI Registers, reducing the burden on industry. Respective AI Offices would determine the information required (e.g., mandatory Measures) and which computed Measures are shared publicly (e.g., via a public-facing website backed by an AI Register). The EU AI Act\footnote{\url{https://eur-lex.europa.eu/eli/reg/2024/1689/oj/eng}} establishes an EU-wide AI database (see Article 71) but only for high risk AI systems and the details have not been developed yet, we believe our proposed framework would be compatible with the EU's requirements while being extensible to include jurisdiction specific requirements. For example, expanding coverage beyond high-risk system coverage. 

By decoupling the core technical framework from specific regulations, we believe work can urgently begin on the proposed framework, which can cater for current and future AI regulation. 

\subsection{Catalog of Standardized Measures and AI Cards}
% public good through AI accountability mechanisms.
\textbf{Catalog of Standardized Measures:} 
%AI accountability practices can make a significant difference even if legal standards and enforceable risk thresholds are introduced solely for the purpose of maintaining transparency among the AI system applications, rather than for gatekeeping that prevents some AI systems from participating in the market.\ntia{1e} The fundamental claim of this proposal is that addressing the tradeoffs between the risks and benefits of complex AI technologies requires a level of transparency that will ensure that AI system consumers can take informed decisions about the choice of AI systems they use, support or invest in.
% To this end, we propose to create a \emph{public registry of AI systems} publishing \textit{standardized evaluation measures}, such as the ones illustrated in Figure~\ref{fig:comparison_card}, and an \textit{AI office} responsible for protecting and promoting AI transparency and accountability.
Measures are defined methods that assess some aspect of an AI system - its definition should ensure results are comparable across AI systems e.g., specifying the data to use during assessment. To this end, we propose to create a \textit{Catalog of Standardized Measures} that specifies which of the Measures are supported and/or required by a given AI Office.   This is compatible with the EU AI Act, which calls for harmonized standards (see Article 40).\footnote{For example, ISO/IEC TR 24027:2021 outlines a set of appropriate methods for the assessment of bias in AI assisted decision making processes}. The EU AI Act promotes collaboration with international partners on metrology, however, it does not yet propose a framework for the modeling, standardization, cataloging and publication of measures as proposed in the framework proposed here. The motivation for standardized measures and examples are discussed in detail in (\S\ref{sec:ai-measures}). The Catalog of Standardized Measures supports versioning to facilitate the addition and enhancement of Measures over time. To reduce the burden on industry, we encourage common Measures across AI Offices, but the decision of requiring a specific Measure to be assessed depends on the regulation implemented by a given AI Office. 

\textbf{Computed Measure:} The result of computing a Measure of an AI system, a computed Measure, can be stored in the AI Register enabling transparent reasoning about performance, fairness and alignment, including AI explainability measures. We advocate for Measures that are specific to intended uses similar to the stratification of requirements in the EU AI Act. These computed Measures of respective AI systems, such as the ones illustrated in Figure~\ref{fig:comparison_card}, would be made available publicly (both human and machine readable) to the extent determined by the respective AI Office. We explain how AI Offices' Assessment Programs support the creation of computed Measures below.

 % can be the result of self assessment tools, third party assessments or verification audits. 
 
 %The requirement would be determined by jurisdiction and may vary by intended use as proposed by the EU AI Act.
% \url{https://www.iso.org/standard/77607.html}

% However, the regulation does not specify whether standardized measures are required, nor does it discuss them in more depth. 
% As the functional specifications of the EU AI database have yet to be drawn up, our proposal may serve as a blueprint.

% See section on EU AI ACT changes and new AI Office tools including registry - https://digital-strategy.ec.europa.eu/en/library/digital-omnibus-ai-regulation-proposal  see second last paragraph  in 1.5.5 - 
%"In addition, the AI Office will explore opportunities to expand the scope of IT tools (currently mostly in development or pre-launch phase) supporting the AI Act to also cover relevant new enforcement activities (i.e. case handling, AI system registry, monitoring and reporting, exchange of information with authorities). "

% other legislations
\textbf{AI Cards for Computed Measures:} 
The representation of computed Measures within the framework extends existing proposals for AI Cards and datasheets with new requirements for
explainability, transparency, and human impact~\cite{golpayegani2024card,GebruDatasheets,googlemodelcardpub}. AI cards are an effective and transparent way of creating a snapshot of an AI system and the computed Measures that it achieves. These would be incorporated into the data model such that AI Offices could mandate that every AI system used in consequential decision-making shall have an AI Card in the AI Register, akin to the EU requiring that every high-risk AI system must deposit its conformity assessment (see Article 43 as well as Article 27 for the fundamental rights impact assessment that must also be conducted prior to deployment) in the EU database before placing the CE mark on its system and deploying it within the EU marketplace. 
% One key difference is the level of detail provided in our proposal compared to the proposed EU conformity assessment. We believe our proposed approach could help shape and form part of a satisfactory conformity assessment. For higher-risk applications, the register maintains AI cards for commercial systems. 
%These AI cards combine existing proposals for transparency, such as model cards \citep{googlemodelcardpub} and data sheets \citep{GebruDatasheets}, %and %also in this proposal have 
%we additionally require
%with new requirements for %information relevant to 
%explainability, transparency, and human impact.
% our proposal
An example of AI cards is Google’s ``model cards''~\cite{googlemodelcard,googlemodelcardpub}. 
AI cards would capture results of an assessment, which could include both model performance and input influence measures (as we discuss in \S~\ref{sec:ai-measures}). The content of AI cards should be readable and entirely structured in an iterative process that includes industry.\ntia{33} That structure should evolve over time, to reflect the developments in the industry and research, e.g., of novel input influence measures. 
AI cards could provide evaluation results for all input features of an AI model and should be flexible enough to accommodate features specific to the given intended use of AI systems.
% and new features that are not a part of a feature ontology. 
% The registry should provide a feature ontology that is specific to the intended use to reflect the specificity of each decision-making problem. 
The structure of AI cards should be easy to use and understand by both humans and machines.

\subsection{Assessment Programs, Automated Assessments and Certificates}
\textbf{Assessment Programs:} AI Offices may provide Assessment Programs, a mechanism for AI Operators to have their AI systems assessed and computed Measures to be stored in the AI Register. Assessment Programs may include software, automated assessment portals, trusted third parties, or assessments directly carried out by the office. Audit firms may be candidates for trusted assessment parties as they carry out similar assessments for security, resilience and financial accounting -- in that case  Measures would need to be well defined to allow safe comparisons across systems and assessors. AI Offices may permit self-assessment where the AI Operator produces the computed Measure. They may also require manual detailed verification assessments where there is sufficient evidence of wrongdoing in the case of high-risk large-scale models. For example, in situations where comparisons of standardized measures suggest that an AI system needs more oversight, a verification assessment could be mandated to verify the suspicion and gather further information. For instance, to verify that the AI system discriminates via proxies, the external auditors would need to observe how the model was trained and whether it learned to use correlated features as proxies for the missing protected attributes, as prior research points out~\cite{grabowicz2022marrying}. 

\textbf{Automated Assessments:} In an effort to lower the burden on the AI Industry, certain Measures could support automated assessments hosted by the AI Office (or their delegates). These can be used to compute the required Measures for AI systems in the AI Register. Automated Assessments typically are performed online as ``QA sessions'' between a given AI model and assessment API after prior authorization. Examples of Automated Assessments include popular LLM benchmarks, such as Artificial Analysis Intelligence Index~\cite{maslej_artificial_2025} or the evaluations for ``Abstract and Reasoning Corpus'' for Artificial General Intelligence (ARC-AGI)~\cite{chollet_arc_2025, chollet_arc-agi-2_2025}. Queries would be posed by the assessment system, and responses generated by the AI system under test. The assessed AI model should specify what features it expects on input. The assessment API will initiate the testing session by providing (synthetic or real-world) data on input and asking the AI model to output decisions or answers for each sample. These assessments have a client-server architecture and characteristics of an arms-length assessment, since the AI Office provides a server, while the AI system’s developer provides a client.  They treat AI models as black-boxes and only require access to their inputs and outputs.  We recommend that Automated Assessments follow the U.S. Defense Advanced Research Projects Agency's (DARPA) SAIL-ON model of AI evaluations, used among others by the ARC-AGI: the AI operators are informed about the goals and structure of an assessment, but not its content. In particular, the AI operators should not know the (test) dataset that they are evaluated on.  Such a dataset could contain novel samples to test an AI system’s resilience, adaptation, and confidence when faced with unexpected real-world inputs.

%\textbf{where should this go - or is it too much detail} While the feature ontology is highly structured, the input to the AI systems can be provided in an unstructured way. For instance, imagine that an AI is evaluating a student CV. That CV could be mapped into a tabular, highly structured, format with semantically meaningful fields, such as race, gender, SAT score. A classifier could be trained on this highly-structured data, but if a LLM is our classifier, then it would not explicitly make use of the tabular representation of the data, but rather use the entire CV as input. However, to compute input influence measures, such as the impact of race on AI system’s decisions, the auditors will still need to extract structured information from CVs and manipulate that structured information to measure how such changes, e.g., changing race and associated information on the CV, impacts LLM's decisions. Such manipulations of inputs to observe AI system outputs correspond to randomized experiments performed by researchers to understand decision-making processes, e.g., the seminal callback studies change the first name on CVs to see whether callback rates from employers will be affected by racial correlations of the name~\cite{bertrand2003are}.

\textbf{Certificates:} Assessing computed Measures can be undertaken voluntarily, and certificates may be issued in relation to AI systems. Each intended use should have a different set of associated certificates; for instance, a certificate that a foundational LLM offers 99\% accurate scientific references, or that an AI system for hiring is not discriminating against race and gender, while achieving top 15\% demographic disparity score in comparison to other AI systems having the same intended use (see respective exemplary hiring AI Cards in ``Register 1'' of Figure~\ref{fig:comparison_card}). Finally, certificates and the aspects they test will evolve over time in order to account for ever-changing industry, research, practice, and the overall AI ecosystem.

\textit{Computed Measures} evaluated by Assessment Programs, can be used to produce \textit{AI cards} and \textit{AI certificates}, which will enable meaningful comparisons across AI systems for specific intended uses.
%
% Another key contribution of this paper is the proposal relating to automated audits and certificates. 
% Within the EU AI Act Article 27 "Fundamental rights impact assessment for high-risk AI systems" (5) it states: "The AI Office shall develop a template for a questionnaire, including through an automated tool, to facilitate deployers in complying with their obligations under this Article in a simplified manner." 
This proposal could support notified bodies tasked with issuing certificates under the EU AI Act (see Article 44), as well as audit reports and the "exit reports" within the AI regulatory sandboxes (see Article 57) that support the completion of the conformity assessment requirement for sandbox participants. 
Article 27 (5) of the AI Act states: "The AI Office shall develop a template for a questionnaire, including through an automated tool, to facilitate deployers in complying with their obligations under this Article in a simplified manner." 
% Moreover, according to Article 58 (2d) access to AI regulatory sandboxes is to be free of charge for SMEs and start-ups to reduce their AI Act regulatory burden.
Thus, our proposal could be of relevance to both certificate-issuing EU notified bodies and the competent authorities tasked with running the AI regulatory sandboxes.
As such, this proposal is well aligned with the AI Act and provides suggestions for the implementation of the AI Act's conformity assessment and the EU's AI database. 

\section{Aligning International Regulation, Industry Burden, and Public Transparency}
\label{sec:alignment}
\begin{table*}[!b]
    \centering
    \begin{minipage}{\textwidth}
    \resizebox{1.0\textwidth}{!}{
    \begin{tabular}{|p{2.5cm}|p{2.5cm}p{2.5cm}p{2.5cm}p{2.5cm}p{2cm}p{2cm}p{2cm}|}
    \hline
        % \textbf{Components proposed by this proposal} & \textbf{\href{https://eur-lex.europa.eu/eli/reg/2024/1689/oj/eng}{EU's AI Act}} & \textbf{\href{https://cset.georgetown.edu/wp-content/uploads/t0592\_china\_ai\_law\_draft\_EN.pdf}{China’s (draft) AI Law}} & \textbf{\href{https://www.gov.uk/government/publications/ai-regulation-a-pro-innovation-approach/white-paper}{United Kingdom}} & \textbf{\href{https://bidenwhitehouse.archives.gov/briefing-room/presidential-actions/2023/10/30/executive-order-on-the-safe-secure-and-trustworthy-development-and-use-of-artificial-intelligence/}{United States (rescinded)}} & \textbf{\href{https://rules.cityofnewyork.us/wp-content/uploads/2023/04/DCWP-NOA-for-Use-of-Automated-Employment-Decisionmaking-Tools-2.pdf}{NYC's Automated Employment Law}} & \textbf{\href{https://ised-isde.canada.ca/site/innovation-better-canada/en/artificial-intelligence-and-data-act-aida-companion-document}{Canada’s AI and Data Act (currently terminated)}} & \textbf{\href{https://techlawpolicy.com/2025/01/the-closing-act-of-2024-south-koreas-ai-basic-act/)}{South Korea’s Basic AI Act}} \\
%  \begin{threeparttable}[b]
  \rowcolor{LightGray}
         \textbf{Jurisdiction}& \textbf{{EU\textsuperscript{1}}}\tnote{1} & \textbf{{China\textsuperscript{2}}}\tnote{2} & \textbf{{UK\textsuperscript{3}}}\tnote{3} & \textbf{{USA\textsuperscript{4}}}\tnote{4} &  \textbf{{NYC\textsuperscript{5}}}\tnote{5} & \textbf{{Canada\textsuperscript{6}}}\tnote{6} & \textbf{{South Korea\textsuperscript{7}}}\tnote{7}\\
    \hline
    \rowcolor{LightGray}
        \textbf{Name} & AI Act & AI Law & A pro-innovation approach to AI regulation & Executive Order on AI & Automated Employment Law 144 & AI and Data Act & Basic AI Act \\
    \hline
    \rowcolor{LightGray}
        \textbf{Status} & Enacted & Draft & White paper & Rescinded & Enacted & Rescinded & Enacted \\
    \hline
        \textbf{AI Office} & ✔ EU AI Office & ❌ & ✔ Office for AI & ✔ Office to coordinate the development of AI planned & ❌ & ✔ AI and Data Commissioner Office & ❌ \\ 
    \hline
        \textbf{AI Register} & ✔ EU Database for High-Risk AI Systems & ✔ National AI supervision platform & ⚫Cross-economy AI risk register & ❌ & ❌ & ❌ & ❌ \\ 
    \hline
        \textbf{Public-facing portal of AI Register} & ⚫ EU Database for High-Risk AI Systems & 〇 & ⚫ AI risk register to be society-wide & ❌ & ❌ & ❌ & ❌ \\ 
    \hline
        \textbf{Self Assessment} & ⚫ Conformity Assessments & ⚫ & 〇 & ⚫ & ❌ & ❌ & ✔ \\ 
    \hline
        \textbf{Automated Assessment} & 〇 & 〇 & ❌ & ❌ & ❌ & ❌ & ❌ \\ 
    \hline
        \textbf{3rd Party Assessment} & ✔ By notified bodies & ✔ When necessary & 〇 & ⚫ & ✔ & ✔ & ❌ \\ 
    \hline
        \textbf{Air-gapped Assessment environments} & ✔ AI Regulatory Sandboxes & ✔ AI supervision and management pilot mechanism & ⚫ AI regulatory sandbox & ⚫ AI testbeds & ❌ & ❌ & ❌ \\ 
    \hline
        \textbf{Standardized Measures} & 〇 & ❌ & ❌ & ⚫ Standardized evaluations of AI systems & ✔ (Disparate) impact ratios & ❌ & ❌ 
        \\ \hline
    \end{tabular}
    }
    \caption{Key components of AI regulations across jurisdictions. Legend: ✔ - planned or implemented, but may be a subset of this proposal; ⚫ - planned, but not specified in comparison to this proposal; 〇 - compatible with this proposal, but not specified; ❌ - not planned. A ``compatible'' component may have been mentioned as a possibility, e.g., in a recital of a secondary legislation, but its implementation currently is not specified, whereas a component that is “not planned” has not been mentioned at all.}
    \vspace*{-5mm}
    \begin{tablenotes}
       \item [1] \scriptsize\ 1. \url{https://eur-lex.europa.eu/eli/reg/2024/1689/oj/eng}
       \item [2] \scriptsize\ 2. \url{https://cset.georgetown.edu/publication/china-ai-law-draft/}
       \item [3] \scriptsize\ 3. \url{https://www.gov.uk/government/publications/ai-regulation-a-pro-innovation-approach/white-paper}
       \item [4] \scriptsize\ 4. \url{https://www.federalregister.gov/documents/2023/11/01/2023-24283/safe-secure-and-trustworthy-development-and-use-of-artificial-intelligence}
       \item [5] \scriptsize\ 5. \url{https://www.nyc.gov/site/dca/about/automated-employment-decision-tools.page}
       \item [6] \scriptsize\ 6. \url{https://ised-isde.canada.ca/site/innovation-better-canada/en/artificial-intelligence-and-data-act-aida-companion-document}
       \item [7] \scriptsize\ 7. \url{https://www.msit.go.kr/eng/bbs/view.do?sCode=eng&mId=4&mPid=2&pageIndex=&bbsSeqNo=42&nttSeqNo=1071}\\
     \end{tablenotes}
%  \end{threeparttable}
    \label{tab:key_comps}
    \end{minipage}
\end{table*}

\subsection{Related AI Regulation and Proposed Policies}
\label{sec:related-regulation}

As AI evolves, regulators have proposed or enacted policies to govern the industry. There is a big variety in the degree of regulation and implementation approach. We argue that fragmented legislation is insufficient, following~\citet{LearnedMiller2020FACIALRT}.
% , who proposed an FDA-like approach to facial recognition technology regulation. 
The passage of local, state, and national bans and moratoria on the use of AI systems---for example, facial recognition in crime prevention or large language models in legal filings---reflects urgent public concern about discrimination, privacy, consent, and surveillance. As covered in the introduction, industry struggles to deal with the complexity, cost, and effort of the regulations they face -- fragmentation only compounds the problem. While we do not believe a single unified global regulation is viable, our proposed distributed AI information sharing framework could help regulators converge on common measures, approaches, and processes while allowing each to capture their own jurisdictional nuances. The burden on industry would be greatly reduced by sharing common terms, identifiers, concepts, and data models in a common distributed framework, much the same as has happened in financial reporting.

% table
Table~\ref{tab:key_comps} compares AI accountability regulations across different jurisdictions. %\footnote{\scriptsize\url{https://oecd.ai/en/dashboards/overview}, \scriptsize\url{https://iapp.org/media/pdf/resource_center/global_ai_law_policy_tracker.pdf}, \scriptsize\url{https://www.techieray.com/GlobalAIRegulationTracker}} 
The AI regulation that is the closest to our proposal is the EU's AI Act. For example, recital 74 in the AI Act calls upon the EC to encourage the development of benchmarks and measurement methodologies for AI systems. 
However, Table~\ref{tab:key_comps} shows that the AI Act does not yet operationalize standardized Measures or Automated Assessments as core components.  
% capAI
While the EU Act envisions Assessments (called ``Conformity Assessments'') and an AI Register (called ``AI Database''), their implementation is not fleshed out yet. The leading research-based proposal for their implementation, known as capAI, is more narrow in scope than our framework.~\cite{floridi2022capai}
% does not envision the use of standardized Measures or Automated Assessments~\cite{floridi2022capai}.
It suggests a conformity assessment consisting of three components: an internal review protocol, a summary datasheet to be deposited in the EU's database, and an external scorecard which can be made available to stakeholders of the AI system, but it does not emphasize the use of standardized Measures or Automated Assessments.
Similarly, a prior proposal for an AI registry, by~\citet{mckernon2024ai}, suggests adding information about “capabilities, architecture, compute used, and security of AI systems”, but it does not propose standardized evaluation to avoid additional burden on AI system developers. 
% However, the propose does “not recommend that a model registry should require AI developers to conduct any particular or standardized evaluation” to avoid additional burden on AI system developers. 
% By contrast, standardized evaluation is key to our proposal. To reduce the burden incurred by entities complying with our proposed AI accountability policy, we propose arms length automated and possibly self assessments.

% In other jurisdictions, the Chinese approach allows for the use of copyrighted material for model training in most cases, and provides intellectual property protections for content created with the assistance of AI technology. 
The UK is adopting an outcomes-oriented approach and has committed to establishing a regulatory sandbox for AI. Canada was the first country in the world to create a national strategy for AI, while Korea was the second country in the world to legislate for AI in 2017. However, what is missing across these different jurisdictions is a unifying framework that efficiently promotes AI transparency. 

In the US, the Executive Order on AI has been rescinded. However, the NIST Artificial Intelligence Risk Management Framework \footnote{https://nvlpubs.nist.gov/nistpubs/ai/nist.ai.100-1.pdf} outlines core concepts of “Govern by Map, Measure, and Manage”. We believe our proposed feature would provide a solid foundation for such an implementation.
% NYC law -- shorten this paragraph?
% By way of another example, see the New York City Local Law 144 on Automated Employment Decision Tools\footnote{\scriptsize\url{https://www.nyc.gov/site/dca/about/automated-employment-decision-tools.page}} for a law that established a mandatory third party annual bias audit that must be published on the deployer's website and updated every 12 months. 
Furthermore, the New York City Local Law 144 on Automated Employment is consistent with our proposal in that it established a mandatory third-party annual bias audit, including certain statistical fairness measures, that must be published on the deployer's website and updated every year. 
However, the law does not facilitate their comparisons across AI systems, does not utilize automated assessments, and has a narrow local scope.
% This bias audit involves deriving and displaying impact ratios which can either be via a selection rate or a scoring rate as follows: (1) the selection rate for a category divided by the selection rate of the most selected category or (2) the scoring rate for a category divided by the scoring rate for the highest scoring category.

Many prior works and policy proposals focus on general-purpose AI (GPAI)~\cite{mckernon2024ai, bengio2025international, jin2023cladder}. Our proposal applies also to other AI systems.
% \footnote{\scriptsize\url{https://hai.stanford.edu/news/holistic-evaluation-of-large-language-models-for-medical-applications}}
Within our proposed framework, GPAI can be treated as one of the intended uses of AI systems, since such models shall be evaluated with special standardized measures (~\citet{bengio2025international, jin2023cladder}). In addition, if GPAI is used for another high-risk intended use, e.g., large language models in medical applications \cite{Clusmann2023}, then respective additional standardized measures might be required.
%\footnote{\scriptsize\url{https://hai.stanford.edu/news/holistic-evaluation-of-large-language-models-for-medical-applications}} 
This is important because most AI systems used in consequential decision-making, e.g., hiring automation or recidivism estimation, are based on custom AI systems that are distinct from GPAI.

Consequently, we are not aware of any work that proposes to introduce a distributed AI information exchange to nudge AI Operators to compete in the space of standardized measures within a lightweight AI accountability framework based on automated assessments and public-facing transparency.

\subsection{Mutual Benefits to Industry and policy-maker by Empowering the Public}
% TODO WB - Reword harvard.
\textbf{Power of the Public:} Today, it is mainly publicly traded companies that develop and market AI products to the public either directly or indirectly through other enterprises. These products include AI consumer services, AI enterprise systems, AI platforms and/or AI software components.  AI’s breakneck speed of innovation is driven by and critically dependent on the support of its shareholders.\footnote{https://fortune.com/2025/12/07/openai-stock-market-risk-sam-altman-alphabet-google-gemini/} AI companies and their shareholders' differences in AI responsibility stances can be so powerful that they challenge the balance of power between directors, managers and the board of directors. ~\cite{Harvard_Law_Review_2025}  Recent evidence from MorningStar shows that shareholders pressure is building as they worry more about company ethics and oversight of the use of AI, than other environment, social and governance (ESG) topics.\footnote{https://www.morningstar.com/sustainable-investing/shareholders-worried-about-ai-use-ethics-proxy-votes-show} One of the most vocal shareholder representative, the fund manager, is heavily influenced by the opinion of the public because they are the ultimate shareholder through the pensions and funds that they manage.  Given trusted information, the public can make informed product and investment decisions that ultimately influence AI companies and their behavior. Therefore, empowering the public with information on comparable Measures and AI cards allows them to exert significant influence over publicly traded AI companies, encouraging responsible AI, perhaps more effectively than wholesale regulation. 

%which sometimes comes close to and the line of a responsibility threshold~\cite{Harvard_Law_Review_2025}.
% , e.g., some companies in relentless search for profit have tried to shield their boards from shareholder influence.\footnote{https://harvardlawreview.org/print/vol-138/amoral-drift-in-ai-corporate-governance/}  
%To encourage responsible AI, this proposal promotes a pragmatic solution that can reduce the friction between industry and regulators while protecting consumers, investors and society. AI companies, like most companies are influenced by shareholders; 
%To date, that threshold has been difficult to define, and regulatory efforts to do so are in their infancy. 

%As we argue below, empowering the public with information on the AI-computed Measures and AI cards can exert significant influence over publicly traded companies, perhaps more effectively than wholesale regulation. 

This approach is not without precedence, public access to the results of AI assessments in the form of computed Measures resemble regulatory transparency mechanisms in other domains. For example, nutrition facts labels required by the FDA, SEC financial disclosures, and energy-efficiency labels all use structured, standardized formats to inform consumers and shape market behavior. Similarly, an AI Office that certifies systems would motivate developers to compete in terms of Measures exposed in the AI Register, much like vehicle safety standards and tests encourage automakers to improve safety features.\ntia{2} By enabling comparisons on \emph{objective} and \emph{transparent} grounds, such an AI Register would positively impact the design of AI systems. 

In the financial sector, ESG scores in Europe have become a critical factor in fund selection amongst savvy investors who want to direct their money away from companies that are ill-governed, socially irresponsible or have environmental profiles that they disapprove of. The EU has mandated the reporting of related scores in the European Sustainability Reporting Standards (ESRS)\footnote{https://finance.ec.europa.eu/news/commission-adopts-european-sustainability-reporting-standards-2023-07-31\_en} to facilitate such market decisions. The EU, via an AI Office, could similarly mandate the mandatory publication of certain predefined standard AI Measures using our proposed framework. Today, without an AI Register with information on AI-centric companies, products, or computed Measures, the public finds it difficult to choose which AI products they want to avoid or which AI firms they want their pension funds to invest in or not.

\textbf{Benefits to the AI Industry:} In other industries, the market power of publicly visible measures is clear from examples such as the various New Car Assessment Programs          (NCAP)\footnote{https://www.euroncap.com/en/about-euro-ncap/}.  NCAP, with support from researchers, encourages car manufacturers to adopt additional new measures and to achieve high assessments in those measures. The car industry responds because they know customers’ money will follow. High-quality manufacturers benefit by helping NCAP develop new measures and showing that their products rate highly. In fact, the industry often goes beyond regulation to innovate new measures for their benefit. For example, it was the motor industry that first brought Anti-Lock Braking System (ABS) to NCAP for measurement, those manufacturers with ABS rated higher than those without. Eventually, regulators mandated ABS for the benefit of all. We would encourage the AI industry to work with AI researchers to develop new Measures that would highlight positive attributes of their AI systems and call out those AI systems that have negative attributes.

Based on parallels with the auto industry and shareholder concerns about AI use, companies that are transparent about their AI products will attract investment. In addition the AI Industry will prefer to collaborate with Assessment Programs, as they may reduce the need for extensive regulation that would impede product development and research. We believe regulators that see the AI industry developing Measures and sharing computed Measures will be less likely to overregulate. In parallel, this has the mutual benefit of empowering the public to influence the direction and areas of focus with the transparency they have from AI Registers. This will allow industry to work together with academic research and policy-makers to develop a glide path to safe, fair and transparent AI while maintaining a strong product development and research direction for the benefit of society. 

%TODO NP: Add line about LLM evaluation awareness test etc
\textbf{Protections Against Gaming Standardized Measures:} Responsible companies will collaborate on the creation of Measures and rogue organizations will quickly be identified for lack of transparency and/or low computed Measures therefore rewarding responsible companies. Inappropriately manipulating computed Measures will be discouraged because of the risk of disclosure and the possible impact on share price and market position. We have seen parallels in the motor industry in relation to VW and emissions tests (\citet{Griffin2018}). As Goodhart's law states, any such metric becomes a target, and then ceases to be useful. 
To mitigate this issue, we propose for assessments using never-before-seen data that is confidential to AI Operators.
%To partially mitigate this issue, we propose to split benchmarking data sets into two similarly distributed subsets, analogous to \emph{test and validation sets} in machine learning. In addition, automated assessments can be conducted in an air-gapped mode, where the tested AI system cannot leak information about the data it is tested on to the external world. Such a technique has been used to test the reasoning capabilities of AI systems in the ARC-AGI semi-private evaluation~\cite{chollet2024openai}. 
%However, if Measures are made on data sets that are unrealistic, these strategies fall short for state-of-the-art AI systems such as large language models. These systems are able to predict that it is being evaluated and will behave differently, e.g., Anthropic's Claude Sonnet 4.5 often exhibits this ``evaluation awareness'' \cite{anthropic2025sonnet}. Therefore, paralleling recommendations from the UK AI Security Institute, we propose that benchmarks should be made realistic and, if evaluation awareness occurs, its frequency should be included as a Measure~\cite{souly2025aisi}.     
We detail mitigation strategies in \S\ref{sec:unseen-data} and Appendix~\ref{sec:prop:bench}.

\textbf{Third Party Independent Assessments:} In the world of ESG, third-party data vendors like Bloomberg publish additional ESG scores about companies in addition to those required by regulation.\footnote{https://assets.bbhub.io/professional/sites/10/Governance-Scores-Fact-sheet.pdf} They do this because those scores are of interest to the market decision makers.  Similarly, our proposed framework would allow third parties to publish their own computed Measures on AI products in AI Registers that permit third-party contributions - e.g. misuse incidents, legal cases etc.
This allows third parties that already publish Measures, such as the UK AI Security Institute (AISI), to formally tie their Measures to AI products and use our framework as a vehicle for propagating results to the public.  

Finally, we caution against overregulation (e.g., mandatory licensing to develop AI technologies) as it could frustrate the development of trustworthy AI, since it would primarily inhibit smaller independent AI system manufacturers from participating in AI development. The EU AI Act is careful to exclude from regulation certain types of scientific academic researchers but does not exclude open-source developers, who are major innovators in the space of accountable AI systems for the public good.
\section{AI Measures: Motivation and Background }
\label{sec:ai-measures}
% We begin by listing the existing barriers and opportunities for effective AI accountability. These barriers can be addressed via regulation that will develop standards and incentives for accountable AI systems.

% \subsection{The Need for Transparency}
% intro to explainability
% As AI systems grow in complexity, it is increasingly difficult to reason about \emph{why} they make certain decisions. Complex predictive algorithms make increasingly high-stakes decisions; however, in order to foster trust in these methods, their underlying decision-making processes need to be better understood. This need gives rise to a rich literature on \emph{explaining} AI systems.

%\subsubsection{AI Transparency}
% Apart from developing appropriate benchmarks, the results of the benchmarks need to be clearly and transparently communicated, explained, and compared across competing  AI systems. 

% \subsection{Motivation}
The proposed framework for AI information exchange enables AI Offices to mandate Measures to be computed within their jurisdiction. In this section we explore the background and motivation behind the type of Measures we envisage. Recent events indicate that there is insufficient transparency and explanation to affected people about the uses, capabilities, and limitations of AI systems.\ntia{3e} 
We illustrate the issue of insufficient transparency using examples and describe how each of these examples can be effectively addressed with specific AI accountability mechanisms (\S\ref{sec:statistical-fairness}-\S\ref{sec:unseen-data}):
\begin{itemize}

    \item 
    Due to suspicions of discrimination in algorithmic hiring, e.g., in AI systems of Amazon and HireVue, the systems were disabled by the AI providers. Reasoning about discrimination may require comparisons of standardized Measures, including statistical (\S\ref{sec:statistical-fairness}) and explainable AI input influence measures (\S\ref{sec:measure-xai}).
    % where U.S. courts make explanation based decisions via the burden shifting framework.
    %The importance of AI system comparisons is particularly evident in the employment sector.
    
    \item Some self-driving cars are claimed to drive larger distances without a crash than humans. These claims are based on measurements, but they do not take into account that humans drive in all conditions, whereas respective AI systems in near-perfect conditions or turn themselves off before a crash, possibly embellishing the statistics of miles driven between crashes \cite{nhsta2022tesla}. To reveal such differences, the standardized Measures shall be appropriate -- such as miles driven in standard conditions -- for a particular intended use of the AI systems (\S\ref{sec:measures-vs-use}) .
    % To account for such difference, there is a need for a greater transparency.
    % Some self-driving systems are designed to turn themselves off seconds before a crash under certain circumstances, allegedly embellishing the statistics of miles driven between crashes. Meanwhile, marketing videos claimed that drivers in these self-driving cars are at the wheel only for legal reasons, implying self-sufficiency. Overall, self-driving systems may be marketed as safer than they are in practice. This practice is officially investigated by the U.S. Department of Transportation \cite{nhsta2022tesla}.
    
    \item OpenAI released a report suggesting that GPT-4 has relatively high IQ, passes a lot of college-level tests, and solves coding problems. Later it was shown that GPT-4 does not achieve as good results on hold-out tests that were outside of the AI training dataset~\cite{narayanan2024ai}. To account for this, AI systems need to be evaluated on data that they have never seen before (\S\ref{sec:unseen-data}).
    % This suggests that high performance to some extent stems from training data contamination; however, this fact received far less attention than the initial report. %LLMs are often presented as means to obtain a professional competitive advantage; this recently led an attorney to use an LLM in an unintended way, risking the attorney's career~\cite{brodkin2023lawyer-a}.
\end{itemize}

Unfortunately, such issues are not addressed by existing AI regulation (\S\ref{sec:related-regulation}). To keep consumers of AI systems informed, and to alleviate such issues, we propose to establish a public AI Register allowing comparisons among different AI systems.
%[TODO: risk also includes \# of people. SM is an example]
In principle, all commercially available AI systems posing a risk to society should be \emph{registered} with the respective AI Office, providing information about their intended use, deployment sector, as well as relevant standardized Measures characterizing the AI system at hand. First, we suggest that standardized Measures depend on the given intended use of AI systems (\S\ref{sec:measures-vs-use}). Then, we identify and discuss the need for different Measures such as statistical (\S\ref{sec:statistical-fairness}) and explainability Measures (\S\ref{sec:measure-xai}). Lastly, we identify challenges to Measure robustness from data exposure and gaming, and propose mitigation strategies (\S\ref{sec:unseen-data}).
% Finally, we note that categorizing risk levels of intended uses of AI systems (\S\ref{sec:risk-levels}) and global convergence or AI accountability policies (\S\ref{sec:global-alignment}) would minimize the burden of AI accountability requirements for AI system providers.
% obligations
% For example, high-risk applications like self-driving cars on public roads need to be heavily regulated regardless of scale, medium-risk applications like facial recognition verification systems would meet heavier obligations as they become more widespread, and low-risk applications like social media platforms would only become regulated when a significant portion of the population uses them.

\subsection{Transparency Measures for an Intended Use}
\label{sec:measures-vs-use}
Standardized Measures need to be considered separately for each intended use of AI systems. 
\begin{itemize}
    \item For hiring automation, we may operate with statistical fairness measures and input influence measures, as the ones shown in Register 1 in Figure~\ref{fig:comparison_card}.
    \item For self-driving AI systems installed in cars, customers may be interested in miles driven without a crash in conditions that are representative of typical human driving conditions.
    \item For LLMs, we may need to use Measures evaluating LLMs on realistic data that has never been released publicly.
\end{itemize}

Additionally, the interpretation of computed Measures depends on intended uses of AI systems.
% The insufficiency of statistical measures of the association between outcomes and protected group identities is apparent also when we look across sectors. 
While in hiring the association of group identity and outcomes suggests unfairness, the same association is perceived differently in the context of health needs across ethnic groups; people of color tend to have worse health status even if we compare individuals having the same healthcare spending, as shown in a recent important work~\cite{obermeyer2019dissecting}. 
The study suggests that admissions to special care programs should be granted to the patients who need them the most, who are more likely to be people of color. 
In this case, the association between outcomes and protected group identities may be judged as fair and justified, opposite to the case discussed in the preceding section.

% For other intended uses of AI systems, there may be a need for entirely different measures than statistical fairness measures. For instance, for self-driving AI systems installed in cars, customers may be interested in miles driven without a crash in conditions that are representative of typical human driving conditions. To achieve such meaningful comparisons, there is a need for specification and  standardization of measures used for each category of intended high-risk use of AI systems.
%

\subsection{Importance of Standardized Measures}
\label{sec:statistical-fairness}

Here, we develop a rationale for using standardized Measures that would allow meaningful comparisons of AI systems, building upon anti-discrimination legislation and practices.\ntia{13}

% from statistical measures of fairness to explainability
% It is difficult to identify a single statistical notion that appropriately evaluates AI fairness, even for specific sectors and narrow well-defined domains.\ntia{10} 
In the private employment sector, there exists well-established anti-discrimination legislation. In the U.S., Title VII of the Civil Rights Act has been in effect since 1964; the federal office that oversees its execution, the Equal Employment Opportunity Commission (EEOC), was established in 1965.
The EEOC uses a statistical notion of fairness, known as \textit{demographic disparity}, measuring the association between hiring outcomes and protected group identities. 
% The EEOC uses a statistical notion of fairness measuring the association between hiring outcomes and protected group identities via the so-called ``80\% rule.'' That rule, however, is solely a rule of thumb. It admits exceptions (e.g., the supreme court case Ricci v. DeStefano~\cite{2009ricci-a}), and U.S. courts instead rely on the burden shifting framework, where the explanations, justifications, comparisons of employment practices, as well as burden shifting, play a central role. 
This Measure is used to determine disparate impact via the well-known ``80\% rule of thumb''. However, U.S. courts rely on the burden-shifting framework, where the explanations, justifications, and comparisons of employment practices play a central role.
In the burden shifting framework, the plaintiff and the employer respond to each other in turns. 
The process is started by the plaintiff pointing out disparate impact, e.g., an association between promotion outcomes and race. This step is referred to as ``disparate-impact liability''.
Then, the employer can provide an explanation (``business necessity'') for this association. 
As a response, the plaintiff can point to an ``alternative employment practice'' that alleviates the disparate impact while achieving the business necessity. 
In this context, explanations can be vague; identifying alternative employment practices is hardly ever possible, since typically plaintiffs and legal clerks lack information about any alternatives.

% \subsubsection{The Importance of Standardized Measures}
% \label{sec:measures-vs-use}

To address this issue, AI accountability mechanisms shall enable comparisons among AI systems. 
We propose that standardized Measures that are important for a given intended use of AI systems are reported in the respective AI Register.
For example, consider three hiring automation systems that have comparable accuracy, but one of them has a significantly higher proportion of demographic disparity, shown in Register 1 in \Cref{fig:comparison_card}. Standardized Measures enable such comparisons between models by highlighting critical differences. With respect to the example in \Cref{fig:comparison_card}, if there exists an alternative hiring procedure that achieves the business necessity (accuracy), but yields lower demographic disparity, then consumers concerned with non-discrimination would prefer System C767. Then, according to Title VII of the U.S. Civil Rights Act, the System A764's procedure is deemed illegal. This exemplifies how comparisons enable market and regulatory incentives -- consumers will choose systems that best aligns with their interests and regulators can observe which systems are abnormally discriminatory. 

% \subsection{New Liability Directives Incentivize AI System Transparency}
Identification of appropriate standardized Measures becomes particularly important as global regulators move in different directions regarding the disparate-impact liability. On one hand, Donald Trump released an Executive Order seeking to ``eliminate the use of disparate-impact liability in all contexts''\footnote{\url{https://www.whitehouse.gov/presidential-actions/2025/04/restoring-equality-of-opportunity-and-meritocracy/}}, including AI systems. 
While disparate impact measures face criticism, it is crucial to consider alternative measures that may be helpful in identifying discrimination in AI systems, such as input influence measures (shown in Figure~\ref{fig:comparison_card}).
On the other hand, the EC has published a proposed directive which would alleviate the burden of proof on EU citizens about the faultiness of an AI system: namely, the adopted Product Liability Directive\footnote{\url{https://www.europarl.europa.eu/legislative-train/theme-a-europe-fit-for-the-digital-age/file-new-product-liability-directive}}. 
% These directives are intended to work with the EU AI Act, which includes Article 86 "Right to explanation of individual decision-making". 
% This article states that "Any affected person subject to a decision which is taken by the deployer on the basis of the output from a high-risk AI system listed in Annex III, with the exception of systems listed under point 2 thereof, and which produces legal effects or similarly significantly affects that person in a way that they consider to have an adverse impact on their health, safety or fundamental rights shall have the right to obtain from the deployer clear and meaningful explanations of the role of the AI system in the decision-making procedure and the main elements of the decision taken." 
This directive shifts the burden of proof towards AI providers. 
In this way, AI providers will be incentivized to compete in the space of civil liability and safety assurance, and to seek sufficient measures proving that their AI systems are compliant and competitive in terms of upholding civil rights such as non-discrimination.

\subsection{Importance of Explainability Measures}
\label{sec:measure-xai}

% intro to explainability
As AI systems grow in complexity, it is increasingly difficult to reason about \emph{why} they make certain decisions. Complex predictive algorithms make increasingly high-stakes decisions; however, in order to foster trust in these methods, their underlying decision-making processes need to be better understood. This need gives rise to a rich literature on \emph{explaining} AI systems.
% intro to explainability
AI explanations measure how each input, or feature, influences the outcome of an AI system. These explanations can be generated using popular off-the-shelf methods such as SHAP and LIME \cite{lundberg2017unified,ribeiro2016why}. Under more complex data such as text and images, these features could be the words from bodies' of text or pixels respectively. 

Prior works %In prior work we
propose using AI explainability measures, such as feature highlighting, to reason about the fairness of AI systems~\cite{grabowicz2022marrying} and to prevent ``fairness gerrymandering'', i.e., a situation where fairness with respect to, say, gender groups may result in unfairness with respect to other groups~\cite{grabowicz2023learning}.\ntia{3} 
% Our work introduces a formal technique for fair and explainable automated decision making that is closely tied to existing anti-discrimination legislation through AI explanations.\ntia{10} 
% The technique inhibits direct discrimination in historical training datasets and enables a restricted use of business necessity attributes that correlate with protected attributes, by preventing their use as proxies for the protected attributes. 
% It achieves this by taking a real-world, possibly discriminatory, decision-making process or AI system (Figure \ref{fig:ii}A) and turning it into a non-discriminatory decision-making AI model (Figure \ref{fig:ii}C), without inducing indirect discrimination through proxies (Figure \ref{fig:ii}B). 
% For instance, in the context of redlining, some financial institutions may use zip code as a way to estimate wealth and probability of paying back the loan if no other financial information is available. 
For instance, to identify that a bank is practicing unjustified redlining, a regulatory office for AI accountability could compare the influence of zip code on loan application outcomes across multiple such institutions. 
If some banks rely on zip codes much more than others, this may correspond to unjustified redlining and these banks can be asked to update their AI systems. 
In summary, this approach ties together concepts of fairness, transparency, and explainability of AI systems.\ntia{6} 
%
%fig 1 here, probably reorder numbers
% \begin{figure*}
%      \centering
%     \includegraphics[width=0.8\textwidth]{assets/ii.png}
%      \caption{Comparisons of input influence values on model outcomes for four different models and three kinds of input features: protected feature (red bars) and features correlated (blue) and uncorrelated (gray) with the protected feature.}
%      \label{fig:ii}
%  \end{figure*}
%
% If we overly focus on the statistical association between outcomes and protected group identities, there will be tradeoffs among different goals of AI accountability, known as ``fairness gerrymandering.'' 
% This refers to instances where statistical fairness with respect to certain protected groups results in statistical unfairness with respect to others, e.g., fairness with respect to gender groups may result in unfairness with respect to ethnic groups.\ntia{3}
% For such cases, prior works proposed to use AI explainability measures~\cite{grabowicz2023learning}. Such explanations are automatically generated and address such tradeoffs. 
% Publicly available explanations for various AI systems would facilitate their collective interpretation and understanding.
We recommend to compute and publish such explanations of AI system outcomes in the AI Register to enable more meaningful comparisons of AI systems, as in \Cref{fig:comparison_card}. 
These accountability mechanisms, combined with sufficient domain expertise would help in determining whether a given AI model discriminates against protected groups~\cite{grabowicz2022marrying}.\ntia{3a}

Some researchers call for creating effective explainable AI models~\cite{rudin2019}. However, approaches based on model-agnostic explanations are less restrictive, as they do not constrain AI system architecture (see Appendix~\ref{sec:prop:exp} for an extensive discussion of explainability measures). 
This difference is crucial, because it enables unconstrained innovation and development of effective AI systems, as well as accurate models of human decisions. 
There exists multiple model-agnostic explanation frameworks, and new ones are under development. 
Choosing the appropriate mode of explanation requires careful consideration. In particular, explanation methods should satisfy provable concepts of fairness, privacy and robustness. 
While these measures are a useful tool for those concepts in AI systems, they can be adversarially manipulated~\cite{slack2020fooling} or provide lower quality explanations to protected subgroups~\cite{balagopalan2022road}. 
Paralleling prior authors’ recommendations, we call for these drawbacks to be considered at the time when assessments and corresponding Measures are designed and developed.

\subsection{Protecting Measure Integrity}
\label{sec:unseen-data}

A limitation of Measures is that they only indicate AI systems performance under conditions matching the data used during assessment.
%Contemporary benchmarks are often narrow in scope, 
Once publicly available, Measures become targets of optimization through mechanisms such as: 
%to such benchmarks when developers tailor their systems to maximize their performance on the benchmark. 
(i) AI systems overfitting to before-seen assessment data, and (ii) deliberate ``gaming'' by AI Operators to tailor their systems to maximize performance on Measures required by AI Offices. 
Compounding these issues, state-of-the-art AI systems can exhibit ``evaluation awareness'' by recognizing that they are being evaluated and altering their behavior accordingly, as demonstrated by Anthropic's Claude Sonnet 4.5~\cite{anthropic2025sonnet}. 
These dynamics produce apparent improvements without sizable gains on previously unseen data or real-world tasks, undermining AI accountability by limiting policy-makers and the public from making informed decisions based on computed Measures. % is exacerbated when the same groups develop benchmarks and AI systems. 
To overcome these limitations, we next propose strategies for developing effective assessments. 
%Effective benchmarks must therefore be realistic and have open-world properties, particularly unexpected inputs, 
%testing not just systems' accuracy, but their ability to identify and adapt to novel data. 

%We propose to split benchmarking data sets into two similarly distributed subsets: (i) publicly available data and (ii) confidential data for air-gapped testing.
% A safe and effective system should report that it is unable to make a decision when insufficient information is provided, or is unable to effectively process the information, rather than simply providing its best guess. 
% This aspect of safety assurance is severely underdeveloped. 
\textbf{Evaluation on Never-Before-Seen Data:} Effective assessments should incorporate open-world properties that test not only AI systems' accuracy, but their ability to identify and adapt to unexpected inputs and novel data. DARPA’s SAIL-ON program recently explored open-world assessments by introducing a gap between AI systems and evaluators, enabling testing on data previously unseen by AI systems. 
This separation allow assessments to more closely emulate real-world deployment scenarios, including their unpredictable ``messiness,'' and evolvement over time, e.g., yearly updates.
%controls the amount of information that AI Operators have about the data and environments in which their AI systems are evaluated.

\textbf{Protections Against Gaming:} 
We propose partitioning assessment data into two similarly distributed subsets, analogous to \emph{test and validation sets} in machine learning, where one subset is confidential. Automated assessments are then conducted in an air-gapped mode that prevents tested AI systems from leaking information about the evaluation data to the external world (see \Cref{sec:prop:bench} for methodology). This approach adds an layer of protection on top using never-before-seen data: it ensures that assessment data cannot be leaked and subsequently utilized to game future assessments. The ARC-AGI semi-private evaluation employs this technique to test AI systems' reasoning capabilities~\cite{chollet2024openai}.  

\textbf{Mitigating and Measuring Evaluation Awareness:}
Even if assessments use never-before-seen data and are air-gapped, AI systems can exhibit evaluation awareness, especially if the assessment design can be easily detectable. Following benchmarks designed by the UK AI Security Institute, assessments should mimic authentic public interactions with AI systems in deployment contexts~\cite{souly2025aisi}. If evaluation awareness occurs, its frequency should be quantified and reported as a Measure, as done in Anthropic's Claude Sonnet 4.5's system card~\cite{anthropic2025sonnet}. 

\section{Conclusion}
Our proposal is synergetic with research in the related areas of responsible, robust, safe, explainable, and interpretable AI, as well as open-world learning. There are many challenges facing regulation in these areas. These include the diversity of policy stance towards AI globally, tensions between industry and regulators over the degree of regulation, lack of consistent Measures of AI systems and the lack of availability of clear information to the public which allows them to make informed decisions. The framework outlined for exchanging information on AI Systems enables industry, researchers and regulators to work together to address these challenges for their mutual benefit.

The public will benefit from the framework by having access to standardized Measures on AI systems that they can trust were created under the supervision of their policy-makers. They can then make informed comparisons of Measures which will drive decisions on which products to use, avoid or place investment. The public have real influence over the strategy of companies, through their role as shareholders, customers and/or users, and having access to trusted information about AI systems empowers them. Policy-makers can leverage and enhance that influence by ensuring the public's access to key Measures are available through the proposed AI Registers. 

In parallel, the proposed framework also benefits regulators, by supporting their different paces of adoption and their diversity in policies, by design. The proposed use of extensible models and federated identifiers similar to those globally accepted in finance will de-risk the technical implementation of their policies. The framework allows for convergence of technical approaches which will encourage cross regulator collaboration while being tolerant of jurisdictional variations.

Finally, a shared federated framework to promote AI transparency and accountability is of benefit to industry. It reduces the  burden of dealing with different technical platforms in each jurisdiction, it allows the sharing of Measures across AI Offices giving them clarity, it supports automated assessments and it allows responsible industry players to work with AI researchers to develop Measures that highlight their responsible efforts while helping to identify rogue operators who may be using AI irresponsibly. 

We believe, with collaboration between industry, research and regulators, the proposed framework will allow the right balance to be struck between levels of innovation, investment and transparency. The framework will be flexible enough to cater not only for the key standardized Measures outlined above but also encourage and support those Measures that need to be developed in the future to keep pace with AI innovations.

\bibliographystyle{plainnat}
\bibliography{zotero,manual_additions}

\appendix
\newpage
\begin{center}
    \textbf{Appendices}
\end{center}

\section{Explainability Metrics}
\label{sec:prop:exp}
%[Individual vs model]
%[company's private data -> model level expl and public data -> individual and model level expl]
%[saliency maps]
%[something about risk levels]

Literature distinguishes between local and global explanations. Thus,
AI registers could publish local and/or explanations for a provided benchmarking dataset. Local explanations provide the influence of features for each individual sample in the dataset, providing an intuitive method to determine if an individual or others similar to them received decisions for the right reasons. Explainability methods may also require context data to generate explanations, e.g., SHAP requires data for integrating out features when measuring influence. Since we do not assume access to proprietary data from AI manufacturers and we call for even comparisons across AI systems, the provided benchmarking dataset should be used as the context data for these methods.  

AI systems could also be required to provide global explanations over entire datasets in the AI register. These are typically aggregates of local explanations over a given dataset and remain consistent with them, e.g., global explanations for SHAP. While not as granular as an individual explanation for a given user/input, global explanations under the same dataset across differing AI systems provides an intuitive method for comparing explanations. Without needing to perform their own aggregations or sample selection, users and experts can observe the differences in the average influence of features across AI systems and either pick the best system for them or contest the system of the manufacturer they are using if its average feature influence is misaligned with other AI systems. %Additionally, if a manufacture desires to release influence measures in the AI registry but has private data, global explanations are appropriate as they do not require information about each sample since it is an aggregation. Conversely, each sample being explained must be provided for local explanations to be interpretable since they compute the influence of each feature with respect to the sample being explained.
Global explanations should not replace local explanations in the AI Register, as they fall short when the influence of features or characteristics on outcomes do not aggregate cleanly. For instance, computer vision explanation methods often use saliency maps to highlight the influence of each pixel of an image \cite{GuidottiSurveyExplaining}. Given that image classification methods are often complex deep learning methods and an image dataset's samples often vary in visual aspects such as perspective and shape of the subject, e.g. the CIFAR-10 dataset of different animals and vehicles \cite{Krizhevsky09learningmultiple}, the aggregation of these saliency maps can result in a map with seemingly random and uninterpretable influence highlights. Therefore, local explanations for individual samples will be needed for interpretable explanations.
%is this understandable? I'm thinking like CIFAR 10. It has a bunch of differently shaped animals and is shot from all sort of angles and distances.

Depending on risk levels and intended use, the office
could also require AI systems to provide explanations and recourse options, which is in line with the EU AI Act Article 86 and accompanying proposed liability directive as outlined above,  to each user and each consequential decision, in addition to including them in the AI register. Similar requirements are suggested in Recital 71~\cite{gdprrecital71} of the EU’s General Data Protection Regulation (GDPR)~\cite{EuropeanParliament2016a}. Recital 71 states that automated decision-making systems should include ``specific information to the data subject and the right to obtain human intervention, to express his or her point of view, to obtain an explanation of the decision reached after such assessment and to challenge the decision.'' 
This Recital, however, is non-binding and provides no technical insight into what type of explanation method should be provided. 
In response, \citet{wachter2017} propose that explanations should provide users with 
\begin{inparaenum}[(1)]
\item understanding on why a particular decision was reached;
\item grounds to contest undesired decisions, and 
\item what could be changed to receive a desired result, i.e., recourse.
\end{inparaenum}
To amend the criticism of the GDPR, \citeauthor{wachter2017} introduces counterfactual explanations. These methods explain  ``how the world would have to be different for a desirable outcome to occur,'' as a means to satisfy these requirements without needing to expose the internal logic of automated decision-making systems. 
For instance, if a loan applicant was denied a loan, a counterfactual explanation will answer 
\begin{inparaenum}[(1)]
    \item what type of applicant was accepted and 
    \item what actions are required to resemble the accepted applicant.
\end{inparaenum} 
Future legislation may incorporate both feature influence and counterfactual explanations, following these proposed principles.

\section{Air-Gapped Assessments}
\label{sec:prop:bench}

To achieve
%our goal of
transparency in AI systems, we propose a benchmarking-based approach to objective model metrics. Benchmarking results are used to compute individual-level explanations, which are reported in full and also in aggregate, alongside other relevant statistics to model performance, such as loss, accuracy, or type 1 and type 2 errors, on data pertaining to various protected groups, as determined by the office on a case-by-case basis.
%We do not establish
% Except in high-risk sectors with legal precedent (e.g., nondiscriminatory employment), 
We do not establish particular standards of what is and is not acceptable, but rather provide benchmarks and assessment results, allowing models to be compared
%to one another
transparently.

Collecting and curating benchmarking datasets is a time-consuming and costly process, however it is necessary to compute explainability metrics, and %establishes a level playing field on which to
allow fair comparisons between models, which we feel is
%a sufficiently valuable objective
sufficiently important so as to justify this cost.
As Goodhart's law states, any such metric becomes a target, and then ceases to be useful. To partially mitigate this issue, we propose to split benchmarking data into two similarly distributed subsets, analogous to \emph{test and validation sets} in machine learning.

The first fold of data is publicly available, and is used in both voluntary self-reported internal assessment and black box automated assessment. We require that modelers not use this benchmarking data in training, % their models,
but this requirement is not directly enforceable, and any attempt to keep this data private would be futile, as it would eventually be fed into commercial systems in benchmarking.
%over the course of standard black-box benchmarking.
Even without explicit intent to game a benchmark, any released benchmarking data is liable to find its way into training data, intentionally or unintentionally, which again would bias results, albeit in subtler manner, and the office must remain aware of this.

%[copy assessment stuff from below]

%\paragraph{External Benchmarking Data Integrity}

Crucially, the second fold benchmarking data, used for external assessment must be %an air-gapped independent fold of data from that made public for voluntary audits and used in internal audits.
remain air-gapped and confidential.
From a security perspective, we must treat any data made public as compromised, as malicious actors could easily use this data to intentionally make better predictions and provide better explanations during voluntary and internal assessments.
We assume an adversarial threat model, and no AI systems is allowed to %communicate
communicate with the outside world in any way after being evaluated on this air-gapped data.
Because %external audit
this benchmarking data must remain confidential, external assessments can only release summary statistics. However, likely even this is unnecessary, as %we can release the
AI cards may report internal assessment summary statistics and individual explanations, and we need only use the external assessments to determine whether further investigation is necessary or rules have otherwise been violated (as
%ideally
Internal and external assessment statistics should match up to a small amount of statistical error.

%[TODO: auditees must check for benchmark data in their own training data, as they can be caught by external audits for this. This should be mentioned earlier?]

%\paragraph{Benchmarking Failure Modes}
Under this setup, we identify several failure modes, all by comparing the results of voluntary internal assessments, automated Black Box assessments, and air-gapped external assessments. Many such errors are explainable by unintentional procedural mistakes or improper data stewardship, but intentional deceit is also a potential explanation, and all such discrepancies should be explored and diagnosed. 
%
%By keeping an air-gapped fold of confidential external assessment data, we gain the ability to detect more forms of wrongdoing.
%We now
In particular, we
assess the internal assessment performance on public benchmarking data, external assessment performance on public benchmarking data, and external assessment performance on air-gapped data.
A discrepancy between internal and external performance on the same dataset indicates some irreproducibility issue or outright fabrication of assessment results, which needs to be investigated.
If these match, but performance on the air-gapped and public datasets do not, this signifies that
%the training data were likely compromised in some way, or some other bad actions were taken. 
either benchmarking data were (ab)used for training, or possibly that modelers intentionally adapted the model to the benchmarking data, likely to gain some unfair advantage in benchmarking.

\section{Support for Researchers}

The AI Accountability regulation would support researchers advising the design and updates to automated assessments and explainability measures used in the AI register. 
% To indirectly support these operations, the government could amplify the allocation of financial resources towards research endeavors in the areas of:
Research areas that are synergetic with the proposed AI accountability policy include
\begin{enumerate}[(1)]
    \item Explainable and interpretable AI methods;
    \item Open-world benchmarks, where there is a gap between model developers and model assessors, such as evaluations in DARPA’s SAIL-ON program;
    \item Collaborative development of ontologies, since the structure of AI cards should be sector-specific and developed collaboratively.
    \item Safe and robust machine learning.
    \item Mechanistic interpretability and causal discovery.
    % \item Identify stakeholders beyond industry and government - coalitions, academic institutes, not for profits, advocacy groups for a holistic approach to accountability.
    \item Acknowledge inherent biases, both systemic and human that when left unchecked create ecosystems that unjustly profile historically disadvantaged communities. \cite{VEubanks2018}
    % \item Social media accountability.
\end{enumerate}

\end{document}